\documentclass[doublecol]{epl2} 
\usepackage{amssymb,amsthm,amsmath,float,epsfig}
\title{Effects of degree-frequency correlations on network synchronization: universality and full phase-locking}
\author{P. S. Skardal\inst{1} \and J. Sun \inst{2,3} \and D. Taylor\inst{1} \and J. G. Restrepo\inst{1} }
\shortauthor{P. S. Skardal, J. Sun, D. Taylor and J. G. Restrepo}
\institute{                    
  \inst{1} Department of Applied Mathematics, University of Colorado, Boulder, Colorado 80309, USA\\
  \inst{2} Department of Mathematics, Clarkson University, Potsdam, NY 13699, USA\\
\inst{3} Department of Physics and Astronomy, Northwestern University, Evanston, IL 60208, USA\\
}

\abstract{
We introduce a model to study the effect of degree-frequency correlations on synchronization in networks of coupled oscillators. Analyzing this model, we find several remarkable characteristics. 
We find a stationary synchronized state that is (i) universal, i.e., the degree of synchrony, as measured by a global order parameter, is independent of network topology, and (ii) fully phase-locked, i.e., all oscillators become simultaneously phase-locked despite having different natural frequencies. 
This state separates qualitatively different behaviors for two other classes of correlations where, respectively, slow and fast oscillators can remain unsynchronized. 
We close by presenting analysis of the dynamics under arbitrary degree-frequency correlations.
}

\begin{document}
\maketitle

%-----------------------------------------------------------------------------------------------------------------------
\section{Introduction} 
%-----------------------------------------------------------------------------------------------------------------------

The research of emergent collective behavior in large ensembles of interacting dynamical systems represents a large and important area of complexity theory~\cite{2003Strogatz,2003Pikovsky,Dorogovtsev2008RMP,Arenas2008PR}. Studying synchronization of coupled oscillators has proven to be particularly useful in modeling complex systems and uncovering generic mechanisms behind synchronization processes. Examples include simultaneous flashing of fireflies~\cite{Buck1988QRB}, cardiac pacemaker cells~\cite{Glass1}, circadian rhythms of mammals~\cite{Yamaguchi2003Science}, collective oscillations of pedestrian bridges~\cite{Strogatz2005Nature}, and chemical oscillators~\cite{Kiss2005PRL}. In many cases, the interactions between oscillators can be described by a complex network. To gain insight into the mechanism behind synchronization, Kuramoto proposed to model the state of each oscillator $n$ by a phase variable $\theta_n$~\cite{Kuramoto1}. When placed on a network, the dynamics of $\theta_n$ is governed by
\begin{equation}\label{eqKuramoto}
\dot{\theta}_n = \omega_n + K\sum_{m=1}^N A_{nm}\sin(\theta_m-\theta_n),
\end{equation}
where $\omega_n$ represents the natural frequency of oscillator $n$, $K$ is the global coupling strength, and $[A_{nm}]$ is the adjacency matrix that encodes the network topology of the underlying system ($n,m=1,2,\dots,N$). 

Although network topology plays a vital role in determining synchronization \cite{Brede2008PLA,Gardenes2011PRL,Restrepo2007PRE,Moreno2004EPL,Sun2009EPL,Ravoori2011PRL,Restrepo2005PRE,Skardal2012PRE,Ichinomiya2004PRE,Pecora1998PRL,Hung2008PRE}, the question of how it influences synchrony is not completely understood. In recent years researchers have started to explore the effect of correlations between oscillator frequency $\omega_n$ and degree $k_n=\sum_{m=1}^{N}A_{nm}$ and observed that in some cases, such correlations can give rise to enhanced synchronizability~\cite{Brede2008PLA} and the emergence of explosive synchronization events~\cite{Gardenes2011PRL}. What, then, is the effect of degree-frequency correlations on synchronization in general? 

We address this question by analytically and numerically studying synchronization in undirected networks (i.e., those for which $A_{nm}=A_{mn}$) with general degree-frequency correlations. These correlations may be characterized by the joint probability distribution of degrees and frequencies $P(k,\omega)$, which we assume to be symmetric about $\omega=0$, i.e., $P(k,-\omega) = P(k,\omega)$. In the classical (uncorrelated) network Kuramoto model \cite{Ichinomiya2004PRE, Restrepo2005PRE,Moreno2004EPL}, the frequencies and degrees are chosen independently, so that the distribution $P(k,\omega)$ can be written as a product of the frequency and degree distributions, $P(k,\omega) = P(k)g(\omega)$. In this Letter we propose a framework to study synchronization in the general case and present detailed results for the case in which the joint distribution is given by $P(k, \omega) = P(k)[\delta(\omega - \alpha k^\beta) + \delta(\omega + \alpha k^\beta)]/2$, i.e.,
\begin{equation}\label{eqCorr}
\omega_n = \pm\alpha k_n^\beta,
\end{equation}
where $\alpha,\beta$ characterize the correlation and the positive and negative signs are chosen with equal probability to maintain zero mean frequency as $N \to \infty$. This particular form of $P(k,\omega)$ is chosen as an illustrative example and is closely related to a model studied numerically in Ref.~\cite{Gardenes2011PRL}  (see their footnote [24]).

This simple model can be used for analyzing the influence of degree-frequency correlations on the synchronization of coupled oscillators and exhibits rich dynamics. We note that $\alpha$ can be scaled out of eqs.~(\ref{eqKuramoto}) and (\ref{eqCorr}) by letting $t\mapsto t/\alpha$ and $K\mapsto\alpha K$. We will therefore use $\alpha=1$ in all figures presented in this Letter. Thus, the free parameters are the coupling strength $K$, the adjacency matrix $[A_{nm}]$, and the correlation exponent $\beta$. We will consider positive correlations and refer to $\beta=1$, $\beta<1$, and $\beta>1$ as linear, sub-linear, and super-linear correlations, respectively.

To measure the degree of synchrony, we introduce the following order parameters. The local order parameter $r_n$ for oscillator $n$, which quantifies the degree of synchrony among the neighbors of node $n$, is defined by $r_ne^{i\psi_n}=\sum_{m}A_{nm}e^{i\theta_m}$, where $\psi_n$ is the local mean phase. The global order parameter is defined by $R=N^{-1}\sum_{n}\frac{r_n}{k_n}$ and measures the degree of synchrony over the entire network.

%\revision{Although degree-frequency correlations can be very complicated, the system defined by eqs.~\ref{eqKuramoto} and \ref{eqCorr} is a simple model that can be used for analyzing the influence of degree-frequency correlations on the synchronization of coupled oscillators with only a few parameters and exhibits rich dynamics. The rest of this Letter is organized as follows. First, we describe the nature of solutions and the remarkable dynamics that arise. Next, we provide analysis for synchronized solutions. We then generalize our results to account for arbitrary correlations $P(k,\omega)$ and finally conclude with a short discussion.}

\section{Description of solutions}
\begin{figure}[t]
\centering
\addtolength{\belowcaptionskip}{-0mm}
\onefigure[width=\linewidth]{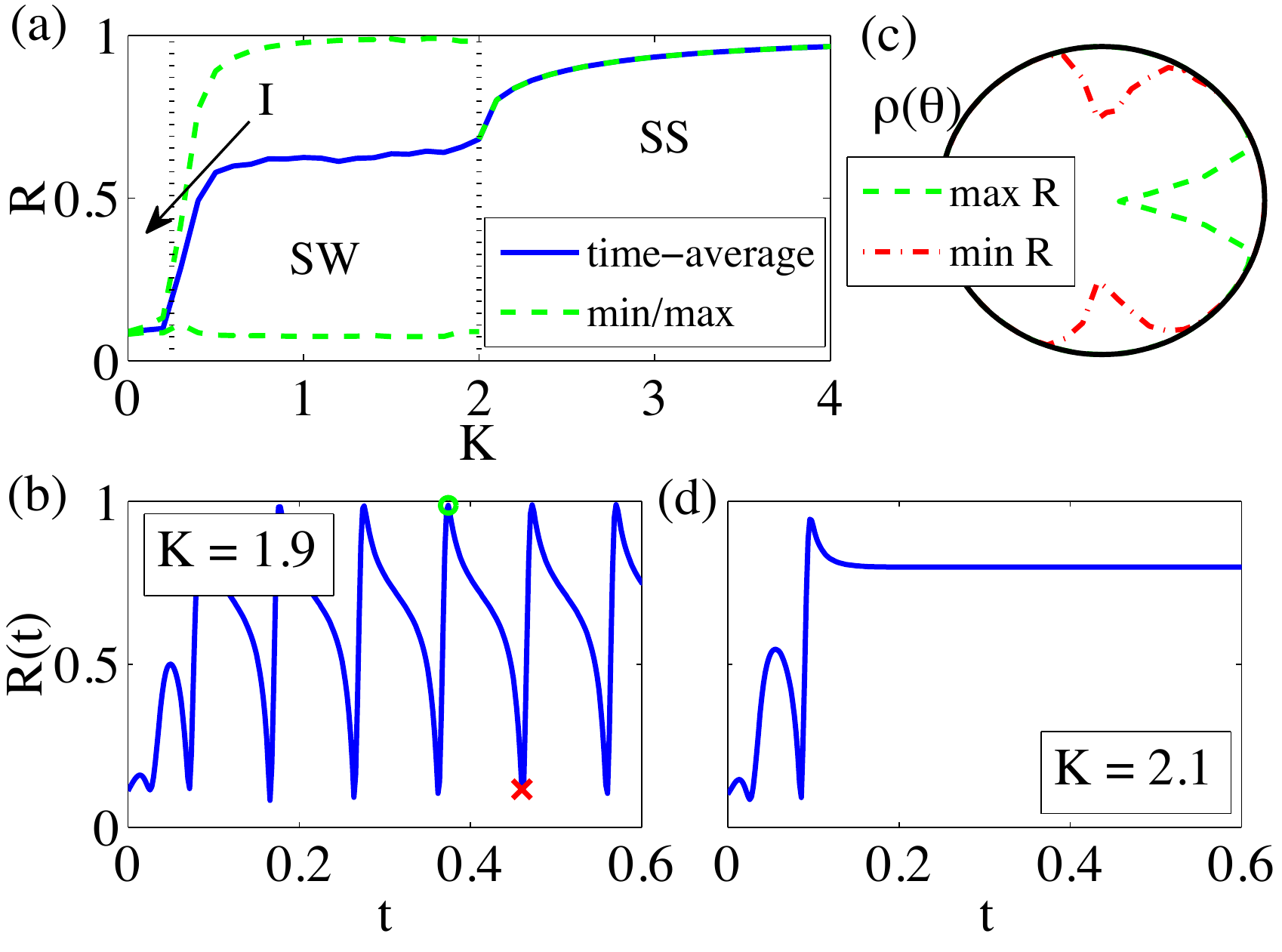}
\caption{(Colour online) Transition from incoherence to coherence for an ER network with parameters $N=1000$, $p=0.1$, and $\beta=1$. (a) Time-averaged (solid blue) and minimum/maximum (dashed green) $R$ versus $K$. (b) Time series $R(t)$ for a SW solution using $K=1.9$. (c) Distribution of phases $\rho(\theta)$ of phases at maximum (dashed green) and minimum (dot-dashed red) $R(t)$ values [times denoted by the green circle and red cross in (b)]. (d) Time series $R(t)$ for a SS solution using $K=2.1$.} \label{example}
\end{figure}

We now briefly describe the dynamics of the steady-state behavior of the system defined by eqs.~(\ref{eqKuramoto}) and (\ref{eqCorr}). We begin by describingthe degree of synchrony as the coupling strength $K$ is varied. In fig.~\ref{example} we plot data from simulations on an Erd\H{o}s-R\'{e}nyi (ER) network \cite{Erdos1960} of size $N=1000$ with link probability $p=0.1$, using a  correlation exponent $\beta=1$. Figure~\ref{example}(a) shows that as the coupling strength $K$ increases, the time-averaged order parameter $R$ also increases towards the value of $1$, as expected. Notably, this $R$-$K$ curve exhibits two transitions, one at the critical coupling strength $K=K_1\approx 0.2$, and the other one at the critical coupling strength $K=K_2\approx 2$ (indicated with vertical dotted lines). This is in sharp contrast to the usual $R$-$K$ curves where a single transition is observed~\cite{Restrepo2005PRE}.

As shown in fig.~\ref{example}(a), these two critical coupling strengths separate three regimes which we denote as incoherent~(I), standing wave~(SW), and stationary synchronized~(SS) states. For $K<K_1$, $R\approx0$ and the system is incoherent, consisting of oscillators that drift independently. For $K_1<K<K_2$, the network exhibits SW solutions characterized by the emergence of two synchronized clusters traveling with opposite angular velocities. Such SW solutions result in the oscillating behavior of $R(t)$, shown in fig.~\ref{example}(b). The distribution of phases $\rho(\theta)$ corresponding to the maximal and minimal $R(t)$ values [e.g., as indicated by the green circle and red cross in fig.~\ref{example}(b), respectively, and shown as dashed lines in fig.~\revision{1}(a)] are depicted in fig.~\ref{example}(c). Note that $R(t)$ achieves its maximum when the distributions of phases for the two clusters overlap (dashed green) and achieves its minimum when they lie on opposite sides of the unit circle (dot-dashed red). Finally, for $K>K_2$ an SS state emerges, yielding a time-invariant $R(t)\approx1$ [fig.~\ref{example}(d)]. 

\begin{figure}[t]
\centering
\addtolength{\belowcaptionskip}{-0mm}
\onefigure[width=\linewidth]{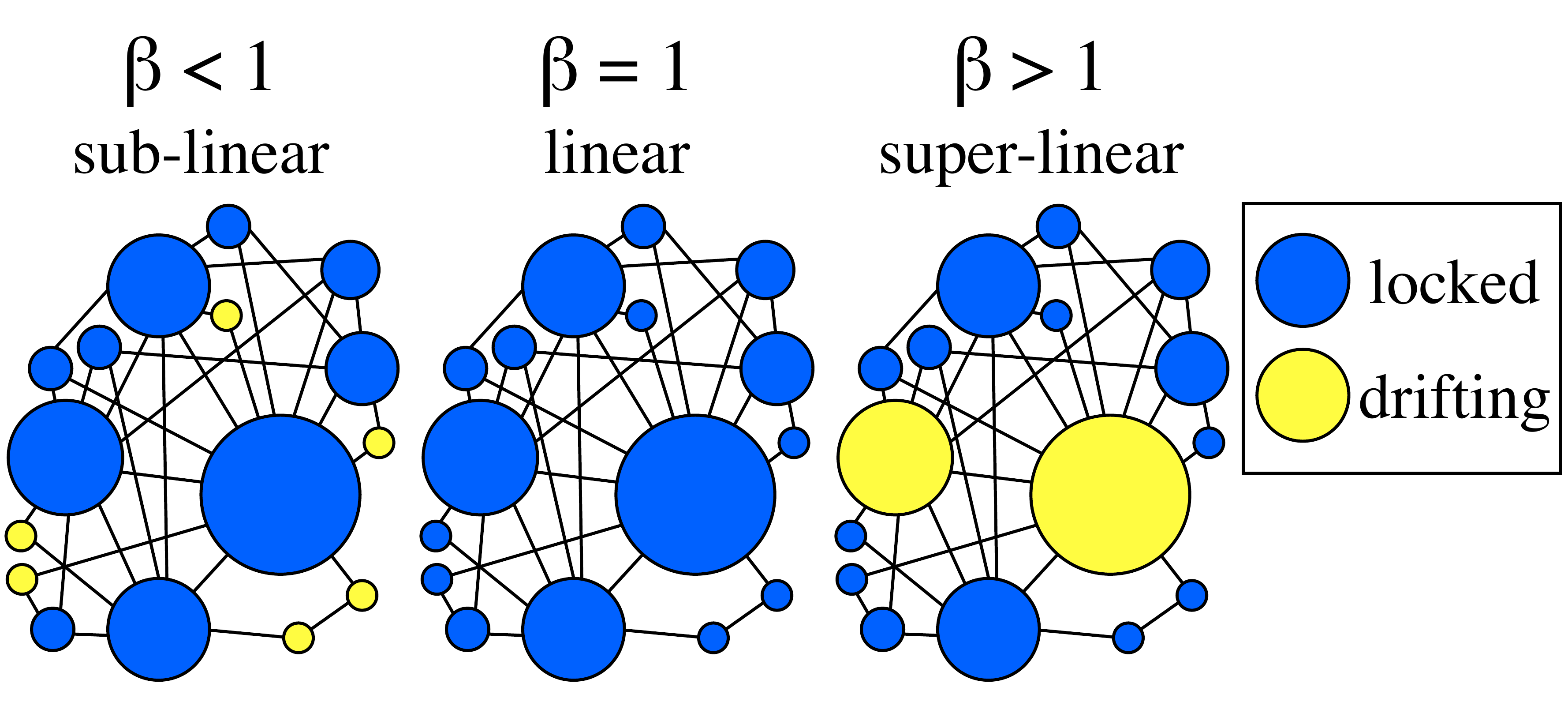}
\caption{(Colour online) Illustration of phase-locking for sub-linear, linear, and super-linear correlations in a network of size $N=16$. Circle radii are proportional to degrees with locked and drifting oscillators colored blue and yellow, respectively. Simulation parameter values are $\beta=0.8,1,1.2$ and $K=1.35,2.1,2.5$, respectively.} \label{cartoon}
\end{figure}

The SS state exhibits remarkable characteristics. In particular, as we will see, with a linear correlation the critical coupling strength for the onset of global synchronization is $K_2=2\alpha$, a universal value that is independent of detailed network topology. The steady-state degree of global synchrony $R$ as a function of $K$ also turns out to be universal in the case of a linear correlation. In sharp contrast, when there is no degree-frequency correlation or when such a correlation is nonlinear, network structure plays a vital role in determining both the critical coupling strength and degree of global synchrony $R$ \cite{Ichinomiya2004PRE, Restrepo2005PRE,Moreno2004EPL}. Furthermore, in the absence of a degree-frequency correlation, only a fraction of the oscillators become phase-locked. The oscillators that are not phase-locked drift indefinitely and typically have either low degrees or high frequencies \cite{Restrepo2005PRE}. However, for a linear degree-frequency correlation ($\beta=1$), whenever the system exhibits global synchrony, all oscillators are locked, which we refer to as full phase-locking. For nonlinear correlations, we find (through both analytical and numerical approaches) that when the correlation is super-linear (sub-linear), drifting oscillators typically exist and are those with high (low) degrees. A linear correlation thus represents a perfect balance between each oscillator's topological (degree) and dynamical (frequency) properties. We illustrate this in fig.~\ref{cartoon}, where we show locked (blue) and drifting (yellow) oscillators from real simulations of a network of size $N=16$ for sub-linear, linear, and super-linear correlations (left to right). Note that for the sub-linear correlation only oscillators with small degrees ($k_n=2$) drift, while for the super-linear correlation only oscillators with large degrees ($k_n\ge8$) drift. The case of a linear correlation corresponds to full phase-locking.

\section{Standing wave solution}
The existence of the SW state can be understood by noting that the frequency distribution of the oscillators is bimodal, a property that has been previously shown to produce SW states for systems lacking degree-frequency correlations \cite{Martens2009PRE}. For example, when $\alpha,\beta=1$ in eq.~(\ref{eqCorr}), the frequency distribution $g(\omega)$ is simply the mirror-reflected version of the degree distribution, $g(\omega)=[P(-\omega)+P(\omega)]/2$. Thus, a unimodal $P(k)$ [in the case of an Erd\H{o}s-R\'{e}nyi network, peaked at $k=p(N-1)$] naturally gives rise to a bimodal $g(\omega)$, which is expected to lead to a SW solution when the separation between the two peaks of $g(\omega)$ is large enough compared to the width of the distribution~\cite{Martens2009PRE}. 

To begin the analysis of the SW solution, we will analyze separately the degree of synchrony in the clusters of oscillators with positive and negative frequencies. To this end, we introduce positive/negative local and global order parameters $r_n^{\pm}e^{i\psi_n^\pm}=\sum_{\omega_m \gtrless0}A_{nm}e^{i\theta_m}$ and $R^\pm=N^{-1}\sum_{n}r_n^\pm/k_n^\pm$, where $k_n^\pm$ is the sum of link strengths connecting oscillator $n$ to oscillators with positive/negative frequencies, $k_n^\pm=\sum_{\omega_m\gtrless0}A_{nm}$. Using the modified local order parameters, eq.~(\ref{eqKuramoto}) can be rewritten as
\begin{equation}\label{eqSW1}
\dot{\theta}_n = \omega_n+K[r_n^+\sin(\psi_n^+-\theta_n)+r_n^-\sin(\psi_n^--\theta_n)].
\end{equation}
We now assume that synchronized oscillators are divided, according to the sign of their frequency $\omega_n$, into two clusters that rotate in opposite directions with angular velocity $\pm\Omega$, so that $\psi_n^{\pm} = \pm \Omega t$. Assuming $\omega_n > 0$ and moving to a rotating frame of coordinates, we define $\phi_n = \theta_n - \Omega t$, and obtain
\begin{equation}\label{eqSW12}
\dot{\phi}_n = (\omega_n - \Omega) - Kr_n^+\sin(\phi_n)-Kr_n^-\sin(\phi_n + 2\Omega t).
\end{equation}
For $\Omega$ not too small, the last term in this equation oscillates rapidly around zero compared to the first two terms and can therefore be approximately averaged out. (Later we will discuss when the value of $\Omega$ we find in our analysis is consistent with this assumption.) We will now look for a solution in which the values of the local order parameters $r_n^+$ are approximately time-independent. We note that this occurs when oscillator degrees $k_n^+\approx k_n/2$ are large enough that fluctuations may be neglected (see~\cite{Restrepo2005PRE} for a discussion). Accordingly, we neglect the last term in eq.~(\ref{eqSW12}), take $r_n^+$ to be independent of time, and find that oscillator $n$ locks with the positive cluster if $|\omega_n-\Omega|\le Kr_n^+$, in which case we have that $\sin(\phi_n) = \frac{\omega_n-\Omega}{Kr_n^+}$; otherwise, it drifts indefinitely. Due to the symmetry of the frequency distribution, drifting oscillators (as a whole) do not contribute to the degree of local or global synchrony~\cite{Restrepo2005PRE}, allowing us to rewrite the local order parameter as
\begin{align}\label{eqSW2}
r_n^+ = \sum_{\substack{\omega_n>0,\\|\omega_n-\Omega|\le Kr_m^+}}A_{nm}e^{i\phi_n}.
\end{align}
Now, since exactly $k_n^+$ terms contribute to the order parameter $r_n^+$, we propose that $r_n^+$ is proportional to $k_n^+$. This approximation has been validated numerically for this and other network-coupled oscillator systems, but is expected to break down for small $r_n^+$ in very heterogeneous networks, e.g., networks with a scale-free (SF) degree distribution $P(k)\propto k^{-\gamma}$ with $\gamma\le2.5$~\cite{Restrepo2005PRE,Ichinomiya2004PRE}. Therefore, we expect the following theory to be valid only for relatively homogeneous networks. Given the definition of $R^+$, we set $r_n^+=R^+k_n^+$. Recalling that $\omega_n = \alpha k_n^{\beta}$ for $\omega_n>0$, we separate eq.~(\ref{eqSW2}) into its real and imaginary part to obtain self-consistent expressions for $R^+$ and $\Omega$,
\begin{align}
R^+ &= \frac{\langle k\rangle^{-1}}{N}\sum_{2|\alpha k_m^\beta-\Omega|\le KR^+k_m} k_m\sqrt{1-\frac{4(\alpha k_m^\beta-\Omega)^2}{(KR^+k_m)^2}},\label{eqSW3} \\
\Omega &= \alpha\frac{\sum_{2|\alpha k_m^\beta-\Omega|\le KR^+k_m}k_m^\beta}{\sum_{2|\alpha k_m^\beta-\Omega|\le KR^+k_m}},\label{eqSW4}
\end{align}
where $\langle k\rangle=\sum_n k_n/N$ and we have also used $k_n^+\approx k_m/2$. For large $N$, eqs.~(\ref{eqSW3}) and (\ref{eqSW4}) can be approximated by
\begin{align}
R^+ &= \langle k\rangle^{-1}\int_{2|\alpha k^\beta-\Omega|\le KR^+k}P(k)k\sqrt{1-\frac{4(\alpha k^\beta-\Omega)^2}{(KR^+k)^2}}dk,\label{eqSW5}\\
\Omega &= \alpha\frac{\int_{2|\alpha k^\beta-\Omega|\le KR^+k}P(k)k^\beta dk}{\int_{2|\alpha k^\beta-\Omega|\le KR^+k}P(k)dk}.\label{eqSW6}
\end{align}
A similar argument would show that $R^-$ satisfies eq.~(\ref{eqSW5}). Eqs.~(\ref{eqSW5}) and (\ref{eqSW6}) give the degree of synchrony in each cluster and must be solved self-consistently. In general eqs.~(\ref{eqSW5}) and (\ref{eqSW6}) need to be solved numerically.

In the case of $\beta=1$, it is possible to find analytically the critical value $K_1$ corresponding to the onset of the SW solution. To do this, we substitute $z=2(\alpha k-\Omega)/KR^+k$ in eq.~(\ref{eqSW5}) and let $R^+\to0^+$, obtaining a critical coupling strength of $K_1=4\alpha^3\langle k\rangle/\pi\Omega_1^2P(\Omega_1/\alpha)$, where $\Omega_1$ is the group angular velocity at onset. If $P(k)$ is unimodal and has a peak at an intermediate $k$ value, e.g. for an ER network, then expanding eq.~(\ref{eqSW6}) about $R^+=0$ yields the condition $P'(\Omega_1/\alpha)=0$. For an ER network with mean degree $\langle k\rangle=(N-1)p$, this yields $\Omega_1=\alpha\langle k\rangle$, $K_1=4\alpha/\pi\langle k\rangle P(\langle k\rangle)$. For a monotonically-decreasing distribution $P(k)$ with minimum degree $k_0$, e.g., a SF network with minimum degree $k_0$, it can be shown that $\Omega_1=\alpha k_0$, which yields a critical coupling strength of $K_1=4\alpha\langle k\rangle/\pi k_0^2P(k_0)$.
We note that, at onset, the period of oscillation of the last term in eq.~(\ref{eqSW12}) is $\pi/\Omega_1$. On the other hand, the timescale of evolution associated with the first two terms is $2\pi/(\omega_n - \Omega_1)$. Therefore, to neglect the last term in eq.~(\ref{eqSW12}) we require $2\Omega_1 \gg \omega_n - \Omega_1$. For a distribution peaked at $k = \hat k$ we require, using $\Omega_1 = \alpha \hat k$ and $\omega_n = \alpha k_n$, that $2\hat k \gg k_n - \hat k$. Therefore we strictly require $2 \hat k \gg \max_n (k_n - \hat k)$. A somewhat less restrictive requirement, which guarantees the condition is valid for most of the oscillators, is $2 \hat k \gg \mbox{rms}(k_n - \hat k)$. In any case, our theory for the onset of the standing wave solution is restricted to networks with a homogeneous degree distribution (e.g. not SF networks).

\begin{figure}[t]
\centering
\addtolength{\belowcaptionskip}{-0mm}
\onefigure[width=\linewidth]{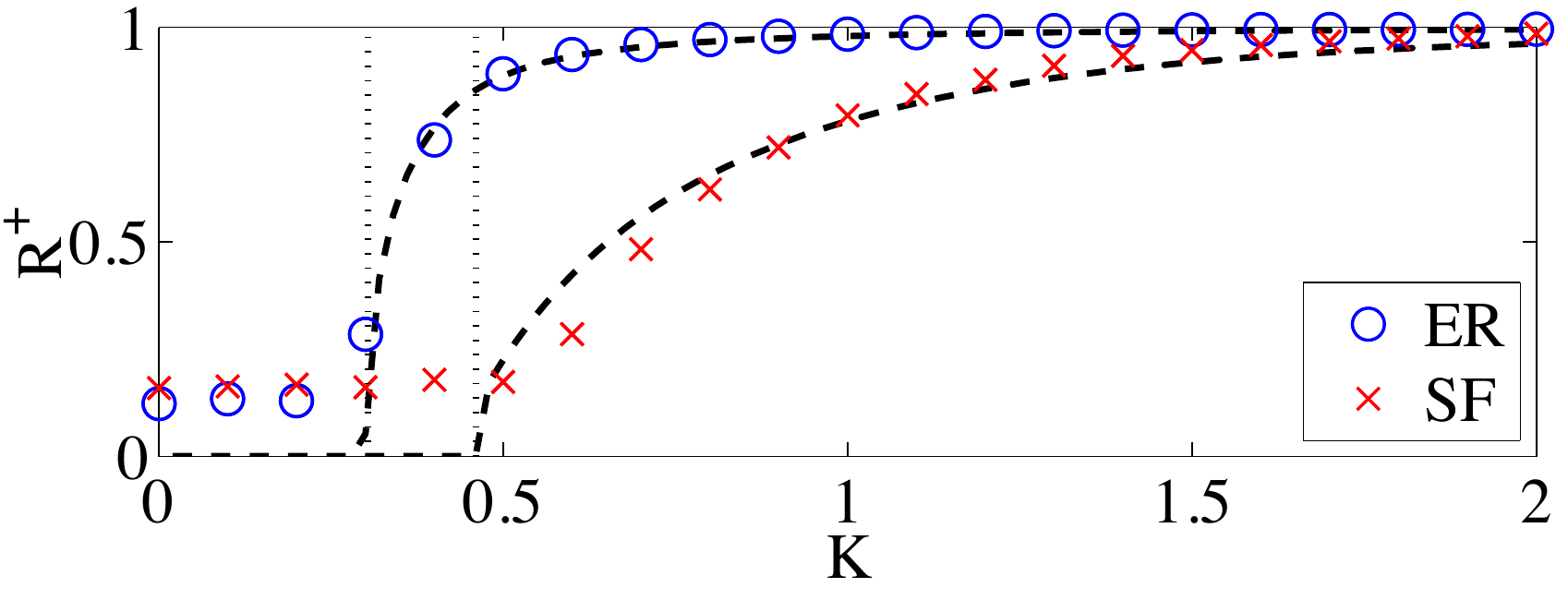}
\caption{(Colour online) Degree of synchrony within positive clusters $R^+$ versus coupling strength $K$ for an ER network with $p=0.1$ (blue circles) and a SF network with $\gamma=5.0$ and $k_0=50$ (red crosses), both of size $N=1000$. Theoretical predictions for $R^+$ given by eqs.~(\ref{eqSW5}) and (\ref{eqSW6}) are plotted in dashed black. Critical coupling strengths $K_1$ are marked with vertical dotted lines.} \label{SW}
\end{figure}

We numerically verify these results by simulating eqs.~(\ref{eqKuramoto}) and (\ref{eqCorr}) with $\beta=1$ over a range of $K$ for an ER network with $p=0.1$ and a SF network with $\gamma=5$ and $k_0=50$ (all SF networks we use in this Letter were generated using the configuration model~\cite{configurationModel}). Both networks are of size $N=1000$. Resulting $R^+$ for the ER and SF networks are plotted in blue circles and red crosses, respectively, in fig.~\ref{SW}. Corresponding $R^-$ values were indistinguishable from $R^+$. Theoretical predictions obtained by solving eqs.~(\ref{eqSW5}) and (\ref{eqSW6}) are plotted as dashed black curves. Critical values $K_1$ for each network are indicated by vertical dotted lines. Results from simulations on the ER network are predicted well by our theory. While our theory is not expected to apply to SF networks, we find reasonable agreement for the SF network with $\gamma=5$. The agreement does break down for smaller values of $\gamma$ (not shown).

\section{Stationary synchronized solution}
We now present an analysis of the SS solution. Using the definition of the local order parameters we rewrite eq.~(\ref{eqKuramoto}) as
\begin{equation}\label{eqKuramoto2}
\dot{\theta}_n=\omega_n+Kr_n\sin(\psi_n-\theta_n).
\end{equation}
We now look for solutions where (i) the synchronized cluster has zero mean frequency and (ii) local order parameters $r_n$ are approximately time-invariant. Oscillator $n$ then becomes phase-locked if $|\omega_n|\le Kr_n$, in which case $\sin(\theta_n-\psi_n)=\omega_n/Kr_n$; otherwise it drifts indefinitely. Due to the symmetry of the frequency distribution, drifting oscillators (as a whole) do not contribute to the degree of local or global synchrony~\cite{Restrepo2005PRE}, allowing us to rewrite the local order parameter as
\begin{equation}\label{eqOrd1}
r_n=\sum_{|\omega_m|\le Kr_m}A_{nm}e^{i(\theta_m-\psi_n)}.
\end{equation}

We now look for solutions that satisfy the following conditions. First, assuming a single synchronized cluster, we set $\psi_n=\psi_m$ for all $n,m$~\cite{Restrepo2005PRE}. We note that this assumption tends to break down when network structure is strongly modular~\cite{Skardal2012PRE}. Second, as in the analysis of the SW solution, since exactly $k_n$ terms contribute to the order parameter $r_n$, we propose that $r_n$ is proportional to the degree $k_n$, i.e. $r_n=Rk_n$. We note that this holds extremely well even for very heterogeneous networks because for SS solutions $r_n/k_n\approx1$. Under these two assumptions, eq.~(\ref{eqOrd1}) becomes
\begin{equation}\label{eqOrd2}
Rk_n = \sum_{|\omega_m|\le KRk_m}A_{nm}\sqrt{1-\left(\frac{\omega_m}{KRk_m}\right)^2}.
\end{equation}

For the linear correlation ($\beta=1$) the dependence of both the summation condition and the square-root term on $k_m$ (and $\omega_m$) vanishes. Looking for the synchronized state, we sum eq.~(\ref{eqOrd2}) over all nodes and find that, after some simplification,
\begin{equation}\label{eqRlin}
R = \sqrt{\frac{1\pm\sqrt{1-4\frac{\alpha^2}{K^2}}}{2}},
\end{equation}
where the $+$ ($-$) sign represents a stable (unstable) solution (numerically determined). This branch of stationary synchronized solutions appears at $K_2=2\alpha$ in the form of a saddle-node bifurcation. Note that in eq.~(\ref{eqOrd2}), since the square-root term becomes constant, the remaining $\sum_mA_{nm}$ term, which encodes the network topology, reduces to the degree $k_n$ which is balanced by the left side of eq.~(\ref{eqOrd2}). Thus, the degree of global synchrony given by eq.~(\ref{eqRlin}) and the critical coupling constant $K_2=2\alpha$ at which the SS solution appears are independent of the detailed structure of the network, which we refer to as universality. This surprising result is found to hold even for networks with degree-degree correlations.

\begin{figure}[t]
\centering
\addtolength{\belowcaptionskip}{-0mm}
\onefigure[width=\linewidth]{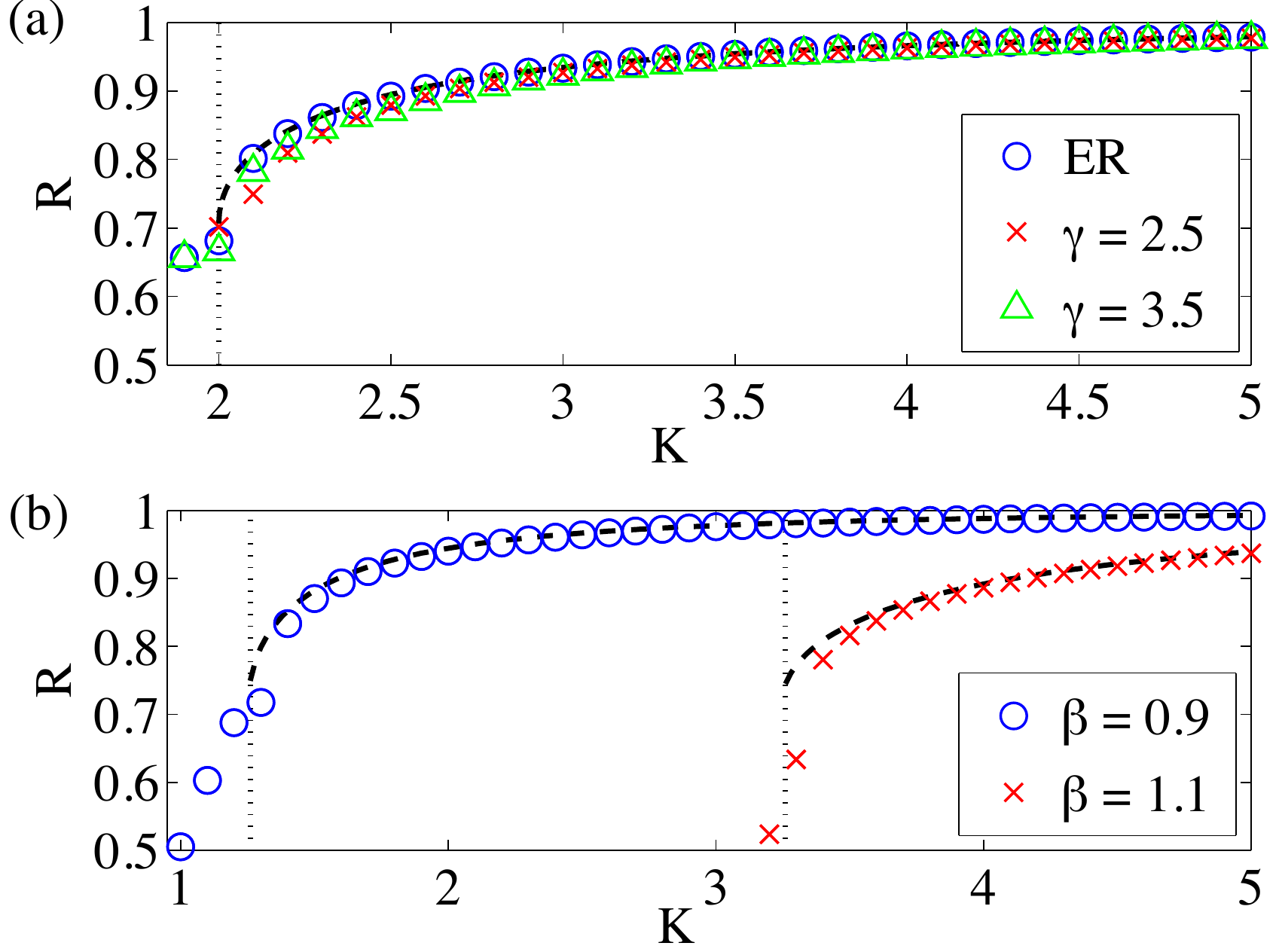}
\caption{(Colour online) Degree of global synchrony $R$ versus coupling strength $K$. (a) Several networks with linear correlations, $\beta=1$. Networks used are ER with $p=0.1$ (blue circles), and SF with $\gamma=2.5$ (red crosses) and $3.5$ (green triangles), both with $k_0=10$, all of size $N=1000$. Theoretical prediction given by eq.~(\ref{eqRlin}) in dashed black. (b) Nonlinear correlations $\beta = 0.9$ (blue circles) and $\beta = 1.1$ (red crosses) on a SF network with $\gamma=3$, $k_0=50$. Theoretical predictions given by eq.~(\ref{eqRnlinint}) in dashed black.} \label{RK}
\end{figure}

We numerically verify these results by simulating eqs.~(\ref{eqKuramoto}) and (\ref{eqCorr}) with $\beta=1$ over a range of $K$ for an ER network with $p=0.1$ and two SF networks with $\gamma = 2.5$ and $3.5$ and $k_0=10$. All networks are of size $N=1000$. Resulting $R$ values for the ER network and SF networks with $\gamma = 2.5$ and $3.5$ are plotted in blue circles, red crosses, and green triangles, respectively, in fig.~\ref{RK}(a). The theoretical prediction given by eq.~(\ref{eqRlin}) is plotted in dashed black. The critical coupling strength $K_2=2\alpha$ is indicated by the vertical dotted line. Results from simulations are predicted very well by our theory, confirming that the detailed network topology is not necessary to describe $K_2$ and $R$ in the SS state for linear correlations. We note that, as opposed to our theory for the SW solution, here we do not need to assume a homogeneous degree distribution.

For nonlinear correlations ($\beta\neq1$), eqs.~(\ref{eqCorr}) and (\ref{eqOrd2}) yield, after summing over $n$,
\begin{equation}\label{eqRnlinsum}
R = \frac{\langle k\rangle^{-1}}{N}\sum_{\alpha k_m^\beta\le KRk_m}k_m\sqrt{1-\left(\frac{\alpha k_m^\beta}{KRk_m}\right)^2},
\end{equation}
where $\langle k\rangle=\sum_{n=1}^Nk_n/N$. For large $N$, we can approximate eq.~(\ref{eqRnlinsum}) with the integral
\begin{equation}\label{eqRnlinint}
R = \langle k\rangle^{-1}\int_{\alpha k^\beta\le KRk}P(k)k\sqrt{1-\left(\frac{\alpha k^\beta}{KRk}\right)^2}dk.
\end{equation}
In general, eq.~(\ref{eqRnlinint}) needs to be solved numerically. 

The critical coupling strength $K_2$ where the stationary synchronized solution is born can be found by solving eq.~(\ref{eqRnlinint}) for the minimum $K$ value where $R>0$ is a solution. Recall that for $\beta=1$ we have $K_2=2\alpha$, which is a universal value independent of the network topology. Increasing (decreasing) $\beta$ effectively spreads (contracts) the set of natural frequencies, therefore impeding (promoting) synchrony and increasing (decreasing) $K_2$. 
%For networks with a sharply peaked unimodal degree distribution (e.g. Erd\H{o}s-R\'{e}nyi) we may use the approximation $P(k)\approx \delta\left(k-\langle k\rangle\right)$ to obtain from Eq.~(\ref{eqRnlinint}) that $K_2\approx2\alpha\langle k\rangle^{\beta-1}$. Note that this approximation recovers $K_2=2\alpha$ for $\beta = 1$.

We numerically verify these results by simulating eqs.~(\ref{eqKuramoto}) and (\ref{eqCorr}) with $\beta\ne1$ on a SF network with $\gamma = 3$ and minimum degree $k_0=50$. Resulting $R$ for $\beta = 0.9$ and $1.1$ are plotted as blue circles and red crosses, respectively, in fig.~\ref{RK}(b). Theoretical predictions for $R$ and the critical coupling strength $K_2$, both obtained by solving eq.~(\ref{eqRnlinint}), are plotted as dashed black and vertical dotted black curves. Results from simulations are predicted very well by our theory. For networks which violate our assumptions by having smaller minimum degrees, e.g., $k_0=10$, we found that $K_2$ as observed from simulations is slightly smaller (larger) for $\beta<1$ ($\beta>1$) than those predicted by eq.~(\ref{eqRnlinint}) (simulations not shown).

\begin{figure}[t]
\centering
\addtolength{\belowcaptionskip}{-0mm}
\onefigure[width=\linewidth]{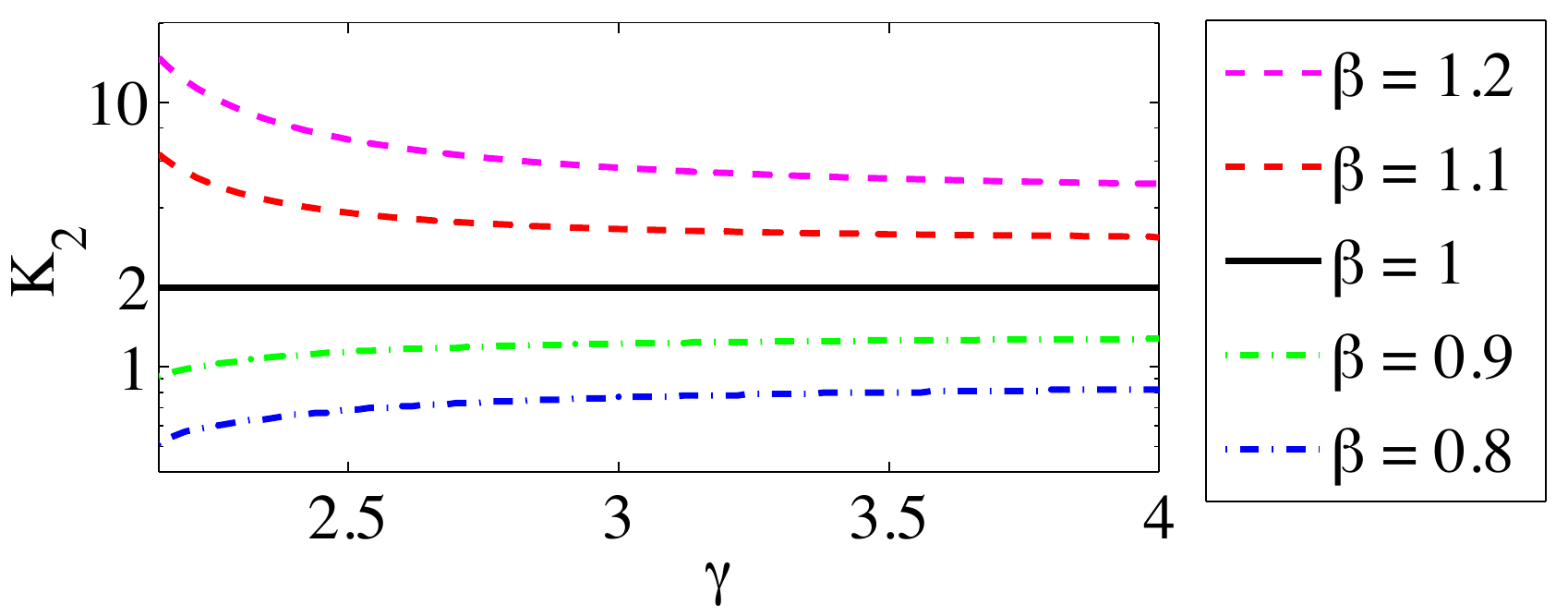}
\caption{(Colour online) Critical coupling strength $K_2$ obtained from eq.~(\ref{eqRnlinint}) as a function of $\gamma$ for SF networks with $k_0=50$ for several values of $\beta$. 
} \label{Kc}
\end{figure}

To further explore the dependence of $K_2$ on network characteristics, we consider SF networks and numerically solve eq.~(\ref{eqRnlinint}) to find $K_2$ given a correlation exponent $\beta$ and degree exponent $\gamma$. Setting the minimum degree $k_0=50$, we plot $K_2$ as a function of $\gamma$ in fig.~\ref{Kc} for increasing values of $\beta\in[0.8,1.2]$, from bottom to top. We see that for $\beta<1$, we have $K_2<2\alpha$, and for $\beta>1$, we have $K_2 >2\alpha$. As the networks become more heterogeneous (i.e., $\gamma$ decreases) $K_2$ curves upward (downward) for $\beta>1$ ($\beta<1$), while $K_2=2\alpha$ remains constant for $\beta=1$.

Having analyzed the SS state, we finally revisit the novel phase-locking behavior introduced in fig.~\ref{cartoon}. Recall our observation that the linear ($\beta=1$) correlation produces full phase-locking, implying that there are no drifting oscillators. In fact this was observed to be a critical case separating the contrasting phase-locking behaviors of sub-linear and super-linear correlations, for which there exist drifting oscillators with low and high degrees, respectively. This interesting  phenomenon can be explained by the locking criterion $\alpha k^{\beta-1}\le KR$ in eq.~(\ref{eqRnlinsum}), assuming that  $K>K_2$. For super-linear correlations ($\beta>1$), oscillators with degree $k\le(\frac{KR}{\alpha})^\frac{1}{\beta-1}$ become locked, while oscillators with high degree and frequency drift, a scenario similar to what has been observed in previous work \cite{Restrepo2005PRE}. For sub-linear correlations ($\beta<1$), the phase-locked population consists of oscillators with degree $k\ge(\frac{\alpha}{KR})^\frac{1}{1-\beta}$, thus leaving oscillators with low degree and frequency drifting. These two qualitatively different behaviors are separated by the critical case of linear correlations ($\beta=1$) for which the dependence on $k$ disappears and the oscillators either all drift or all phase-lock. While we have not performed rigorous experiments testing these critical locking degrees, the results in fig.~\ref{cartoon} are in agreement with our theory.

\section{General Correlations}
We finalize our analysis by noting that, although in this Letter we focused on a specific form for the degree-frequency correlations [i.e., eq.~(\ref{eqCorr})], in the general case of a joint distribution $P(k,\omega)$ symmetric about $\omega=0$, our analysis still holds and results generalize. For the SS solution, we find that eq.~(\ref{eqRnlinint}) is replaced with
\begin{equation}\label{rgeneral}
R = \langle k\rangle^{-1}\int_0^\infty\int_{|\omega|\le KRk}P(k,\omega)k\sqrt{1-\frac{\omega^2}{(KRk)^2}}d\omega dk.
\end{equation}
For a general distribution $P(k,\omega)$, the SW solution will not appear if the distribution of frequencies is not sufficiently bimodal. Otherwise, we may replace eq.~(\ref{eqSW5}) with
\begin{align}
R^+ &= \langle k\rangle^{-1}\int_0^\infty\int_{2|\omega-\Omega|\le KR^+k}P(k,\omega)k\nonumber\\&\hskip25mm\times\sqrt{1-\frac{4(\omega-\Omega)^2}{(KR^+k)^2}}d\omega dk.
\end{align}

\section{Conclusion}
In many applications of network-coupled dynamical systems, a central questions is how the dynamics and network structure give rise to emergent collective behavior \cite{Brede2008PLA,Gardenes2011PRL,Restrepo2007PRE,Sun2009EPL,Ravoori2011PRL,Restrepo2005PRE,Skardal2012PRE,Ichinomiya2004PRE,Pecora1998PRL,Moreno2004EPL,Hung2008PRE}. For instance, in many systems the contribution of the network structure is encapsulated in one or more eigenvalues and eigenvectors of the network adjacency~\cite{Restrepo2007PRE} or Laplacian matrices~\cite{Pecora1998PRL,Sun2009EPL,Ravoori2011PRL}. Here we find that if degree-frequency correlations are chosen appropriately, then the network structure has virtually no influence on the resulting synchronization properties. 

Full phase-locking, i.e., the simultaneous entrainment of all oscillators, in heterogeneous oscillator systems is also a novel finding. Typically, an extremely large value of $K$ is needed to entrain all the oscillators in a large network when the oscillators are heterogeneous \cite{Restrepo2005PRE,Dorfler2011SIAM}. However, in the presence of a linear degree-frequency correlation, all oscillators become phase-locked simultaneously as the coupling constant passes the critical value for global synchrony, $K_2$. This unexpected phenomenon emerges despite the presence of strong heterogeneity in both the network structure and oscillator dynamics. Another remarkable observation is that, for sublinear correlations, the locked oscillators are those with a frequency which is most different from the mean.%, albeit those with a frequency close to the mean drift.

In addition to analyzing the case of eq.~(\ref{eqCorr}), we have presented a general formalism to analyze synchronization of network-coupled oscillators with degree-frequency correlations.  This framework may potentially be used to optimize the synchronization properties of networks, which have been recently realized experimentally \cite{Leyva2012PRL}.

Two recent preprints \cite{Sonnenschein2012,Coutinho2012} independently studied additional aspects of degree-frequency correlations.

\acknowledgements
Supported by NSF Grant No.~DMS-0908221 (P.S.S., D.T., and J.G.R.) and ARO Grant No.~61386-EG (J.S.).


\begin{thebibliography}{0}
	\bibitem{2003Strogatz}
		\Name{S. H. Strogatz}
		\Book{Sync: The Emerging Science of Spontaneous Order}
		\Publ{Hyperion}
		\Year{2003} % checked
	\bibitem{2003Pikovsky}
		\Name{A. Pikovsky, M. Rosenblum, \and J. Kurths}
		\Book{Synchronization: A Universal Concept in Nonlinear Sciences}
		\Publ{Cambridge University Press}
		\Year{2003} % checked
	\bibitem{Dorogovtsev2008RMP} 
		\Name{S. N. Dorogovtsev, A. V. Goltsev, \and J. F. F. Mendes}
		\REVIEW{Rev. Mod. Phys}{80}{2008}{1275} % checked
	\bibitem{Arenas2008PR} 
		\Name{A. Arenas \etal}
		\REVIEW{Phys. Rep.}{469}{2008}{93} % checked
	\bibitem{Buck1988QRB} 
		\Name{J. Buck}
		\REVIEW{Q. Rev. Biol.}{63}{1988}{265} % checked
	\bibitem{Glass1} 
		\Name{L. Glass \and M. C. Mackey}
		\Book{From Clocks to Chaos: The Rhythms of Life}
		\Publ{Princeton University Press}
		\Year{1988} % checked
	\bibitem{Yamaguchi2003Science} 
		\Name{S. Yamaguchi \etal}
		\REVIEW{Science}{302}{2003}{1408} % checked
	\bibitem{Strogatz2005Nature} 
		\Name{S. H. Strogatz \etal}
		\REVIEW{Nature (London)}{438}{2005}{43} % checked
	\bibitem{Kiss2005PRL} 
		\Name{I. Z. Kiss, Y. Zhai, \and J. L. Hudson}
		\REVIEW{Phys. Rev. Lett.}{94}{2005}{248301} % checked
	\bibitem{Kuramoto1} 
		\Name{Y. Kuramoto}
		\Book{Chemical Oscillations, Waves, and Turbulence} 
		\Publ{Springer-Verlag}
		\Year{1984} % checked
	\bibitem{Brede2008PLA} 
		\Name{M. Brede}
		\REVIEW{Phys. Lett. A}{372}{2008}{2618} % checked
	\bibitem{Gardenes2011PRL} 
		\Name{J. G\'{o}mez-Garde\~{n}es, S. G\'{o}mez, A. Arenas, \and Y. Moreno}
		\REVIEW{Phys. Rev. Lett.}{106}{2011}{128701} % checked
	%\bibitem{Wang2011PRE} H.~Wang and X. Li, Phys. Rev. E {\bf 83}, 066214 (2011). % checked
	\bibitem{Restrepo2005PRE} 
		\Name{J. G. Restrepo, E. Ott, \and B. R. Hunt}
		\REVIEW{ Phys. Rev. E}{71}{2005}{036151} % checked
	\bibitem{Ichinomiya2004PRE} 
		\Name{T. Ichinomiya}
		\REVIEW{Phys. Rev. E}{70}{2004}{026116} % checked
	\bibitem{Moreno2004EPL} 
		\Name{Y. Moreno \and A. F. Pacheco}
		\REVIEW{Europhys. Lett.}{68}{2004}{603} % checked
	\bibitem{Restrepo2007PRE} 
		\Name{J. G. Restrepo, E. Ott, \and B. R. Hunt}
		\REVIEW{Phys. Rev. E}{76}{2007}{056119} % checked
	\bibitem{Skardal2012PRE} 
		\Name{P. S. Skardal \and J. G. Restrepo}
		\REVIEW{Phys. Rev. E}{85}{2012}{016208} % checked
	\bibitem{Pecora1998PRL} 
		\Name{L. M. Pecora \and T. L. Carroll}
		\REVIEW{Phys. Rev. Lett.}{80}{1998}{2109} % checked
	\bibitem{Sun2009EPL} 
		\Name{J. Sun, E. M. Bollt, \and T. Nishikawa}
		\REVIEW{Europhys. Lett.}{85}{2009}{60011} % checked
	\bibitem{Ravoori2011PRL}
		\Name{B. Ravoori \etal}
		\REVIEW{Phys. Rev. Lett.}{107}{2011}{034102} % checked
	\bibitem{Hung2008PRE}
		\Name{Y.-C. Hung \etal}
		\REVIEW{Phys. Rev. E}{77}{2008}{016202} % checked
	\bibitem{Erdos1960} 
		\Name{P. Erd\H{o}s \and A. R\'{e}nyi}
		\REVIEW{Pub. of the Math. Inst. of the Hung. Acad. of Sci.}{5}{1960}{17} % checked
	\bibitem{Martens2009PRE} 
		\Name{E. A. Martens \etal}
		\REVIEW{Phys. Rev. E}{79}{2009}{026204} % checked
	\bibitem{configurationModel} 
		\Name{A. Bekessy, P. Bekessy, \and J. Komlos} 
		\REVIEW{Stud. Sci. Math. Hung.}{7}{1972}{343}
	\bibitem{Dorfler2011SIAM}
		\Name{F. D\"{o}rfler \and F. Bullo}
		\REVIEW{SIAM J. Appl. Dyn. Syst.}{10}{2011}{1070}
	\bibitem{Leyva2012PRL} 
		\Name{I. Leyva et al.}
		\REVIEW{Phys. Rev. Lett.}{108}{2012}{168702} % checked
	\bibitem{Sonnenschein2012} 
		\Name{B. Sonnenschein, F. Sagu\'es, \and L. Schimansky-Geier}
		\REVIEW{preprint}{}{2012}{arXiv:1208.6491}{}
	\bibitem{Coutinho2012} 
		\Name{B. C. Coutinho \etal}
		\REVIEW{preprint}{}{2012}{arXiv:1211.5690v2}{}
\end{thebibliography}
\end{document}